\documentclass[12pt]{article}
\def\ra{\rightarrow} \def\sw{Schwarzschild}

\newcommand{\ee}{\end{equation}}    \newcommand{\be}{\begin{equation}}
\newcommand{\eeq}{\end{equation}}    \newcommand{\beq}{\begin{equation}}
\def\ba{\begin{eqnarray}}  \def\ea{\end{eqnarray}}
 \def\Si{\Sigma} \def\t{\tau} \def\r{\rho} 
\def\f{\phi}    \def\th{\theta}  \def\D{\Delta}
     
\def\a{\alpha} \def\b{\beta}  \def\G{\Gamma} \def\de{\delta}
   \def\e{\epsilon} 
\def\Th{\Theta}  \def\2{\frac{1}{2}}  
\def\ba{\begin{eqnarray}}  \def\ea{\end{eqnarray}}
\def\2{{1\over2}}

\begin{document}
\title{Integrating the geodesic equations\\
in the Schwarzschild and Kerr space-times\\
using Beltrami's ``geometrical" method}
\author{
Dino Boccaletti\footnote{Dipartimento di Matematica, Universit\`a di
Roma \lq\lq La Sapienza", Roma, Italy \protect \\
e-mail boccaletti@uniroma1.it },
Francesco Catoni, Roberto Cannata\footnote{
ENEA, C.R. Casaccia, Roma, Italy \protect\\
e-mail cannata@casaccia.enea.it},
Paolo Zampetti\footnote{
ENEA, C.R. Casaccia, Roma, Italy \protect\\
e-mail zampetti@casaccia.enea.it }}
\date{\today}
\maketitle
{\bfseries ABSTRACT} - We revisit a little known theorem due to Beltrami,
through which the integration of the geodesic equations
of a curved manifold is accomplished by  a method which, even if
inspired by the Hamilton-Jacobi method, is purely geometric.
The application of this theorem to the Schwarzschild and Kerr metrics
leads straightforwardly to the general solution
of their geodesic equations.
This way of dealing with the problem is, in our opinion, very much in keeping
with the geometric spirit of general relativity.
In fact, thanks to this theorem we can integrate the geodesic
equations by a geometrical method and then verify that
the classical conservation laws follow from these equations.\\

\tableofcontents

\section{Introduction}

In the preface 
to his
classic 
textbook ``The Analytical Foundations
of Celestial Mechanics" \cite{Win}, A. Wintner wrote in 1941
\lq\lq\ldots 
even the classical literature of the great century of Celestial
Mechanics appears to be saturated with
rediscoveries\ldots" 
He then illustrated this remark with some examples of results that had been rediscovered
more than once, retracing their history back to the first discoverer. 
This is also true of
the field of differential geometry and especially of its applications to
general relativity (GR), where Malcolm MacCallum has used the
particularly appropriate term \lq\lq literature horizon"
to describe the loss of familiarity with results that accompanies each new generation
of relativists. 
We propose that an old theorem on geodesics
due to
Eugenio Beltrami (1835--1900)  \cite{Bel} , which as far as we know has not
been quoted in
texts on 
differential geometry since the publication of the
classic 
books
by L. Bianchi \cite{Bi} and L.P. Eisenhart \cite{ei}, deserves to be
rediscovered. In his theorem Beltrami showed that the geodesic
equations can be
integrated through a method substantially analogous to the method of
separation of variables used in the integration of the Hamilton-Jacobi
equation.

In GR we have some important spacetimes which are exact solutions 
of the Einstein equations and whose metric tensor components are known explicitly
in a given system of coordinates. Starting from these components, one can write
the geodesic equations\footnote{For the geodesic equations (\ref{eul}) and the formulas
applied below, we refer to the standard texts on general relativity.
In particular, in the following we shall refer to two of the most
widespread texts \cite{MTW, dino}.}
\beq\label{eul}
\frac{d^2 x^l}{ds^2}+ \G^l_{ik}\,\frac{dx^i}{ds}\,\frac{dx^k}{ds}=0, \label{geod}
\ee
and then try to integrate them to determine the paths of test particles.
The \sw\ spacetime (whose timelike geodesics can be used to calculate
the advance of the perihelion of Mercury) and the Kerr metric (representing the
gravitational field outside a rotating body or of a mathematical black hole)
are two important examples whose geodesics can yield important physical results. 

Two methods are typically used to integrate the geodesic equations. 
Either one starts with the Lagrangian equations of motion 
(obtained
from 
a Lagrangian ${\cal L}$ given by ${\cal L}=g_{ik}
\,(dx^i/ds)\,(dx^k/ds)$, where for timelike geodesics $s$ may be
identified with the proper time) or with the corresponding Hamilton-Jacobi equations,
in  both cases representing a mechanical system governed only by a kinetic energy term,
in which the effects of the gravitational
field are represented by the curvature of the spacetime associated with the metric
which determines this kinetic energy function.
Now one is dealing with a mechanical system again instead of pure geometry.
In our eyes, this approach seems to be a step backwards with respect to the spirit of GR.
The motion of a test particle in a gravitational field	is interpreted as
the motion of a free particle in a curved spacetime
which turns out to follow a geodesic.
On the other hand, 
a completely ``geometric" integration of the geodesic equations
can be performed without referring to the equivalent point particle mechanical system.
Once the geometric problem has been solved, the constants
of 
integration can be interpreted as physical constants that are the
first integrals of the motion in the classic approach.
For its historical 
interest we reproduce in Sect. \ref{dcpgeo} a concise translation of
the relevant section of Beltrami's article, only updating its notation.
\section{Beltrami's integration methods for geodesic equations}
\label{dcpgeo}
We recall Beltrami's method for obtaining the solutions of geodesic equations
 \cite[Vol.~I, p.~366]{Bel}, an extension
of the Hamilton-Jacobi integration method used for
integrating the equations of motion in dynamics.
This method originating with Beltrami in the later 1800s does not appear in recent books \cite{doC};
the last exposition of this method is reported in Eisenhart's  classic book \cite[p.~59]{ei}.\\
Let us consider an $n$-dimensional semi-Riemannian manifold $V_n$ whose
metric is represented by
\beq
d\,s^2=g_{i\,h}dx^i\,dx^h
\qquad (i,\;h= 1,2,...n)\,.
\label{B1}
\eeq
If $U,\;V$ are any real functions of the $x^i$ ($i= 1,2,...n$), the invariants defined by
\ba
\D_1\,U=
g^{ih}\frac{\partial\,U}{\partial x^i}\,\frac{\partial\,U}
{\partial x^h}\equiv g^{ih}U_{,i}\,U_{,h}, \label{B2} \\
\D\,(U,\,V)=
g^{ih}\frac{\partial\,U}{\partial x^i}\,\frac{\partial\,V}
{\partial x^h}\equiv g^{ih}U_{,i}\,V_{,h} \label{B2A}
\ea
are called {\it Beltrami,s differential parameters of the first order}.
Since the equations $U=const$, $V=const$ represent ($n-1$)-dimensional
hypersurfaces in $V_n$, $\D_1\,U$ represents the squared length of the
gradient of $U$ as well as of a vector orthogonal to the
hypersurface $U=const$ ; for the same reason, if $\D(U,\,V)=0$, then
the two hypersurfaces $U=const$ and $V=const$ are orthogonal.\\
The Beltrami's theorem states:\\
 Let us consider the equation
\beq
\D_1 \,U=1,   \label{B}
\eeq
the solution of this equation depends on an additive constant and on $N-1$
essential constants $\a_i$ \cite{ei}. Now, if we know a complete solution of
Eq.~(\ref{B}), we can obtain the equations of the geodesics from the following
theorem \cite[p.~299]{Bi}, \cite[p.~59]{ei}:
{\it when a complete solution of Eq.}~(\ref{B}) {\it is known, the equations of the
geodesics are given by
\beq
\frac{\partial U}{\partial \a_i}=\b_i \label{t1} \,,
\eeq
where $\b_i$ are
arbitrary constants, and the geodesic arclength is given by the value of
$U$.\/}

We demonstrate the theorem following Beltrami's article \cite{Bel}: \\ Let us
return to differential parameters (\ref{B2}), (\ref{B2A}). We can associate contravariant
components of the gradient with the covariant ones
\beq
U^{,h}=g^{hi}U_{,i}\,. \label{B5}
\eeq
Since (\ref{B1}) and (\ref{B2}) are reciprocal quadratic forms (since the
coefficients in (\ref{B2}) are reciprocal to the ones in (\ref{B1})),
because of their invariance we can write
\beq
ds^2=\D_1\,U\; dk^2\,,	 \label{B4A}
\eeq
where $dk$ is an infinitesimal scalar quantity introduced to keep
the differential homogeneity. From Eq.~(\ref{B4A}), we have obviously
\beq
dk=\frac{ds}{\sqrt{\D_1\,U}}\,.\label{B4}
\eeq
Moreover, owing to the properties of the reciprocal forms and keeping the
differential homogeneity expressed by Eq.~(\ref{B4}), we can write
\beq
\2\frac{\partial (ds^2)}{\partial (dx^i)}
\equiv g_{ih}dx^h =U_{,i}\, \,dk\,,  \label{B3}
\eeq
which can be rewritten as
\beq
dx^h=g^{ih}\,U_{,i}\, \,dk\equiv U^{,h}\,\, dk \label{B3'}\,.
\eeq
Combining (\ref{B4}) with (\ref{B3'}), we get
\beq
\frac{dx^h}{ds}=\frac{U^{,h}}{\sqrt{\D_1\,U}} \label{B31}\,.
\eeq
By multiplying (\ref{B3}) by $dx^i$ and summing, we shall get
\beq
ds^2=dU\,\,dk \,,	  \label{B32}
\eeq
where $dU$ is the differential of the function $U$.

In an analogous way, if the increments $\de x^i$  are the components of
a vector {\it indicating another direction\/} from that of components
$dx^i$, we shall get by multiplying by $\de x^i$ and summing
\beq
g_{ih}\de x^i\,dx^h=\de U\, \,dk\,.  \label{B3A}
\eeq
It is evident that, if we {\it take the vector indicated by $\de x^i$ tangent to a
hypersurface $U=const$ , we have $\de U=0$ and then the vectors, of components
$dx^i$ and $\de x^i$ respectively, turn out to be orthogonal. Then the
``displacement" $dx$ of (\ref{B3}), is orthogonal to the hypersurface
$U=const$\/}.

Finally, if we eliminate $dk$ from Eqs.~(\ref{B4A}, \ref{B32}), we find
\beq
\D_1\,U=\left(\frac{dU}{ds}\right)^2  \label{B3B}	\,,
\eeq
where $dU$ represents the increment due to a variation $ds$ which, for
what we have seen, is orthogonal to the hypersurface $U=const$.

Let us now consider a curve represented in parametric form as a function
of a parameter $t$ and set
$$
\dot s
=ds/dt\qquad \mbox { and }
\qquad\dot x^i=d x^i/dt\,.
$$
From (\ref{B1}), one has
\beq
\dot s^2=g_{ih}\dot x^i\,\dot x^h\,. \label{B6}
\eeq
For this quadratic form we can repeat the preceding arguments and, in
particular, in place of (\ref{B3}) we shall have
\beq
\2\frac{\partial\, \dot s^2}{\partial\, \dot x^i}
\equiv \dot s\frac{\partial \, \dot s}{\partial \,\dot x^i}
\equiv g_{i\,h} \dot x^h
=U_{,i}\, \, k; \mbox{ for } i=1,...,n \,,\label{B7}
\eeq
where now $k$ is a finite constant of proportionality.

Furthermore, since $\dot s^2$ and $\D_1 U$ are two scalar quantities in
the same metric, we can write
\beq
\dot s^2=\D_1 U\, \,k^2 \label{B8A} \,,
\eeq
from which
\beq
 k=\frac{\dot s}{\sqrt{\D_1 U}}\,. \label{B8}
\eeq
By eliminating $k$ from (\ref{B7}) and (\ref{B8}) we obtain
\beq
\frac{\partial \dot s}{\partial \dot x^i}=\frac{U_{,i}}{\sqrt{\D_1 U}}
\,. \label{B9}
\eeq
The condition for a curve to be a geodesic, as we know, in Lagrangian
form (where ${\cal L}=\dot s$ is the Lagrangian) is given by the $n$ equations
\beq
\frac{\partial\,\dot s}{\partial x^i}
=\frac{d}{dt}\left(\frac{\partial \dot s}{\partial \dot x^i}\right) \label{B10}.
\eeq

Let us assume that we know $n$ first integrals of these equations
\beq
\dot x^i=f^i(x^1,...x^n) \label{B15}
\eeq
and substitute them into $\dot s=\sqrt{g_{ih}\dot x^i\,\dot x^h}$
and ${\partial \dot s}/{\partial \dot x^i}$ of (\ref{B10}) which,
in this way, will turn out in the end to be functions only of the variables
$x^i$. If we indicate by $\left(\frac{d}{dx^i}\right)$ 
the ``total" derivatives with respect to $x^i$ and make explicit the
derivatives of the two sides of (\ref{B10}), we get the equations
\ba
\left(\frac{d\,\dot s}{d\, x^i}\right)
=\frac{\partial\,\dot s}{\partial\, x^i}
+\frac{\partial\,\dot s}{\partial\, \dot x^r}
 \frac{\partial\,\dot x^r}{\partial \,x^i}\,,
\label{B11}\\
\frac{d}{dt}\left(\frac{\partial\dot s}{\partial \dot x^i}\right)
=
\frac{\partial}{\partial\, x^r}\frac{\partial\,\dot s}{\partial\, \dot x^i}\;
\dot x^r \label{B12}\,,
\ea
where the implicit dependence on $x^i$, once one have substituded the first
integrals into the expressions of $\dot s$ and ${\partial \dot s}/{\partial
\dot x^i}$, has been exploited.

Now, if we differentiate with respect to $x^i$ the identity
$\dot s=({\partial \dot s}/{\partial \dot x^r})\;\dot x^r$,
we get
\be
\left(\frac{d\dot s}{dx^i}\right)=\dot x^r\,
\frac{\partial}{\partial x^i}\frac{\partial \dot s}{\partial \dot x^r}+
\frac{\partial \dot s}{\partial \dot x^r} \frac{\partial \dot x^r}
{\partial x^i} \label{B13}\,.
\ee
From a comparison of (\ref{B11}) and (\ref{B13}) it follows that
\beq
\frac{\partial\,\dot s}{\partial\, x^i}
=\frac{\partial}{\partial  x^i} \frac{\partial \dot s}{\partial  \dot x^r}\;\dot x^r \,.
\eeq
These equations and (\ref{B12}) allows us to write the system (\ref{B10}) in the
{\it pfaffian form}
\beq
\left(\frac{\partial}{\partial\, x^r}\frac{\partial\,\dot s}{\partial\,
\dot x^i}-\frac{\partial}{\partial\, x^i}
\frac{\partial\,\dot s}{\partial\, \dot x^r}\right)\;\dot x^r =0\,,
\quad (i=1,...,n)\,.
\label{pfa}
\eeq
The system (\ref{pfa}) is satisfied if it is possible to assign a function $U$
(potential) such that
\beq
\frac{\partial \dot s}{\partial\, \dot x^r}=
\frac{\partial U}{\partial  x^r}\equiv U_{,r}\,. \label{B17}
\eeq
From (\ref{B9}) it follows that this condition is satisfied if $U$
is a solution of Eq. \ref{B}. Then, if we find a solution of the partial differential
equation (\ref{B}),
from (\ref{B31}) we obtain the equations of the geodesics as functions of
$s$ by solving a system of first order ordinary differential  equations.
If (\ref{B}) is satisfied, we have from (\ref{B3B}) that $U\equiv s$ and then
$U$ {\it represents the line element along the geodesics\/}.

On the basis of what we have summarized above, we can finally enunciate the
fundamental result of Beltrami due to which the second integration can
be avoided: {\it If we know a complete solution of the partial differential
equation} (\ref{B}), {\it we can obtain the geodesic equations by
differentiation steps alone.\/}

The demonstration is straightforward.
If (\ref{B}) is satisfied, from (\ref{B3B}) it follows that the length
of the orthogonal segments between two hypersurfaces $U=const$ is the same.
According to a theorem due to Gauss, the lines which cross orthogonally
the fields $U=const$ are geodesic lines and the hypersurfaces $U=const$ are
said to be geodesically parallel.

Let us consider a complete solution of (\ref{B}) which, in addition to an
obvious additive constant, will contain $n-1$ other arbitrary constants $\a_l$.
By differentiating (\ref{B}) with respect to these constants, we obtain
\beq
\frac{\partial \D_1\,U}{\partial \a_l}\equiv 2\,\D\left(U,\,\frac{\partial U}
{\partial \a_l}\right) =0 \mbox{\,, for } l=1,...,n-1\,, \label{B19}
\eeq
which tells us that the hypersurfaces
$V_l \equiv {\partial U}/{\partial \a_l} = const$
and $U=const$ are orthogonal. If we now put
\beq
V_l \equiv \frac{\partial U}{\partial \a_l}=\b_l\,,  \mbox{ for } l=1,...,n-1
\label{B20}\,,
\eeq
the curves of intersection of these hypersurfaces, orthogonal to $U=const$, are
geodesics and have been obtained through differentiation, without being
obliged  to solve the differential equations  (\ref{B31}). \\ The demonstration of
Beltrami's theorem is complete.  \\

Obviously the above theorem is particularly useful for applications when we
have a complete solution of (\ref{B}) at our disposal or this solution is
easily obtained.\\
A general case in which the solution of Eq. (\ref{B}) is obtained in integral form
has been obtained by Bianchi \cite{Bi, ei}\footnote{In this paper we do not apply this result and
obtain the solution of
Eq.~(\ref{B}) directly for metrics that are not in the Liouville form.}: \\
If the fundamental form (\ref{B1}) can be re-expressed	in the
generalized Liouville form
\beq
ds^2=[X_1(x_1)+X_2(x_2)+...+X_N(x_N)]\sum_{i=1}^n e_i (dx^i)^2 \,,\label{Lio}
\eeq
where $e_i=\pm 1$ and
$X_i$ is a function of $x^i$ alone, a complete integral of Eq.~(\ref{B}) is 
\beq
U=c+\sum_{i=1}^n\int\sqrt{e_i(X_i+\a_i)}\,dx^i,     \label{supc01}
\eeq
where $c$ and $\a_i$ are constants, the latter being subject to the condition
$\sum_{i=1}^n \a_i=0$. \\
In this case the geodesic equations (\ref{B20}) are immediately given by
\beq
{{\partial U}\over{\partial \a_l}}\equiv {{1}\over{2}}\int{{e_i\,dx_i}\over{
\sqrt{e_i\,(X_i+\a_l)}}}=\b_l.	 \label{supc02}
\eeq

\section{The geodesic equations for the Schwarzschild and Kerr metrics}
\label{swge}
\subsection{The Schwarzschild metric}
We start from the standard form of the so called \sw \ metric \cite{Sw}
\beq
d\,s^2=\frac{r-\a}{r}\,d\,t^2-\frac{r}{r-\a}\,d\,r^2-
r^2(d\,\th^2+\sin^2\th\,d\,\f^2);\quad \a=2MG \label{Sw}
\eeq
Applied to this metric 
Eq.~(\ref{B}) gives\footnote{We replace the previous symbol $U$ with $\t$,
that recalls the physical meaning of proper time along the geodesics.}
\beq
\frac{r}{r-\a}\left(\frac{\partial\, \t}{\partial \,t}\right)^2
-\frac{r-\a}{r}\left(\frac{\partial\, \t}{\partial \,r}\right)^2
-\frac{1}{r^2}\left[\left(\frac{\partial\, \t}{\partial \,\th}\right)^2
+\sin^{-2}\th\left(\frac{\partial\, \t}{\partial \,\f}\right)^2\right]
=1
\,.
\label{eqge}
\eeq
If we set
\beq \label{tau}
\t=A_1\,t+A_2\,\f+\t_1(r)+\t_2(\th)\,,
\eeq
Eq.~(\ref{eqge}) can be solved by the method of ``separation of variables."
We obtain
\beq \label{varsep}
r^2\left[A_1^2\frac{r}{r-\a}-\frac{r-\a}{r}\left(\frac{d\, \t_1}{d \, r}\right)^2-1\right]
= \left(\frac{d\, \t_2}{d \,\th}\right)^2
  +\frac{A_2^2}{\sin^2\th} \,.
\eeq
Since the right hand side is positive, we can set both sides equal to a
separation constant
$A_3^2$ to obtain
\ba
r^2\left[A_1^2\frac{r}{r-\a}-\frac{r-\a}{r}\left(\frac{d\, \t_1}{d \,
r}\right)^2-1\right]=A_3^2	    ,	       \label{varsep1}	    \\
\left(\frac{d\, \t_2}{d \,\th}\right)^2+\frac{A_2^2}{\sin^2\th} =A_3^2,
\label{varsep2}
\ea
which then gives
\ba
\t_1=\pm\int\frac{\sqrt{A_1^2\,r^4-r\,(r-\a)(r^2+A_3^2)}}{r\,(r-\a)}\,d\,r,
\label{tau1} \\
\t_2=\pm\int\frac{\sqrt{A_3^2\sin^2\th-A_2^2}}{\sin\th}\,d\,\th \,.
\label{tau2}
\ea
If in the right hand sides of (\ref{tau1})  and (\ref{tau2}) we choose the positive sign,
the geodesic equations (\ref{t1}) become
\ba
\frac{\partial \t}{\partial A_1}
\equiv t+A_1\int\frac{r^3\,dr}{(r-\a)\sqrt{A_1^2r^4
-r(r^2+A_3^2)(r-\a)}} =B_1 \,,
\label{geosw1}\\
\frac{\partial \t}{\partial A_2}
\equiv \f-A_2\int\frac{d\th}{\sin\th\sqrt{A_3^2\sin^2 \th-A_2^2}}
\equiv \f + \sin^{-1}[\sinh\e\,\cot\th] = B_2 \,,
\label{geosw2} \\
\frac{\partial \t}{\partial A_3}
\equiv -A_3\int\frac{dr}{\sqrt{A_1^2r^4-r(r^2+A_3^2)(r-\a)}}
+ A_3\int\frac{\sin\th\, d\th}{\sqrt{A_3^2\sin^2\th-A_2^2}}
\nonumber \\
\equiv -A_3\int\frac{dr}{\sqrt{A_1^2r^4-r(r^2+A_3^2)(r-\a)}} -\sin^{-1}[\cosh\e\,\cos\th]
=B_3 \,,
\label{geosw3}
\ea
where in the explicitly evaluated integrals we have set $A_3=A\,\cosh\e,\;A_2=A\,
\sinh\e$.

Eq.~(\ref{geosw2}) is the same as Eq.~(4) on p.~645 of Misner et al.\ \cite{MTW}
which establishes the planar character of the orbit.
In fact, with a suitable
rotation of the angle $\f$ ($\f_0=B_2$), we obtain from (\ref{geosw2}) the result
$\sin\f=-\sinh\e\,\cot\th$ and by substituting this into the defining equations
for polar coordinates, one finds the orbit in $x,\,y,\,z$ space as a function of
the parameter $\th$:
\[
\left\{
\begin{array}{l}
x=r(\th)\,\sin\th\,\cos\f\equiv r(\th)\,\sin\th\,\sqrt{1-\sinh^2\e\,\cot^2\th}
\,,\\
y=r(\th)\,\sin\th\,\sin\f\equiv -r(\th)\,\sinh\e\,\cos\th
\,,\\
z=r(\th)\,\cos\th
\,.
\end{array}
\right.
\]
It immediately follows that $y=-\sinh\e\,z$, i.e., the orbit lies in this plane.
We remark that Misner et al.\ obtain this result by applying
the Hamilton-Jacobi method.

We can now calculate the proper time $\t$ from the identity
\beq
\t-A_1\frac{\partial \t}{\partial A_1}-A_2\frac{\partial \t}{\partial A_2}
-A_3\frac{\partial \t}{\partial A_3}=\t-A_1\,B_1-A_2\,B_2-A_3\,B_3
\label{tau5}
\eeq
obtained by substituting the partial derivatives by the constants $B_i$. In the left
hand side we evaluate $\t$ by using Eq.~(\ref{tau}) in which $\t_1$ and $\t_2$ are
given by Eqs.~(\ref{tau1}) and (\ref{tau2}). We also evaluate the other three terms of
the left hand side by using the first identities of Eqs.~(\ref{geosw1}), (\ref{geosw2}) and
(\ref{geosw3}), respectively. We group the integrals in $d\,r,\;d\,\th$ and see that
the resulting integral in $d\,\th$ vanishes. Finally we obtain $\t$ as a function of $r$
via an elliptic integral:
\beq
 \t=A_1\,B_1+A_2\,B_2+A_3\,B_3
-\int\frac{r^2\,dr}{\sqrt{A_1^2r^4 -r(r^2+A_3^2)(r-\a)}}
 \,.
\label{tempro}
\eeq

At this point we must do two things: compare our results
with the existing literature and at the same time identify the
constants $A_1,\,A_2,\,A_3$ that we have introduced
with the relevant physical constants of the classic approach.
To do this, we set $\th=\pi/2$, which is the a priori value everyone assumes.
Then from (\ref{varsep2}), $A_2^2=A_3^2$ and the sum
of (\ref{geosw2}) and (\ref{geosw3}) we get
\beq
\f=A_3\int\frac{dr}{\sqrt{A_1^2r^4-r(r^2+A_3^2)(r-\a)}} +const \,,
		     \label{geosw4}
\eeq
which can be re-expressed as
\beq
\f=A_3\int\frac{dr}{r^2\,\sqrt{A_1^2-(1+A_3^2/r^2)(1-\a/r)}} +const \,.
		     \label{geosw5}
\ee
In the same way (\ref{geosw1}) can be rewritten in the form
\beq
t=-A_1\int\frac{dr}{(1-\a/r)\,\sqrt{A_1^2-(1+A_3^2/r^2)(1-\a/r)}} +const \,.
\label{geosw6}
\eeq

If we compare our equations (\ref{geosw6}) and (\ref{geosw5}) with (101.4)
and (101.5) of Landau-Lifshitz \cite{LL}, which are obtained starting
from the Hamilton-Jacobi equation, we find that the equations are the same,
apart from the different symbols and units used in them. In the limit $\a/r\ll 1$
(as is the case for planetary motion), (\ref{geosw5}) becomes
\beq
\f=A_3\int\frac{dr}{r^2\,\sqrt{(A_1^2-1)-A_3^2/r^2}} +const
		     \label{geosw7}\,.
\eeq
If we compare (\ref{geosw7}) with the relevant Newtonian equation (see
Boccaletti-Pucacco \cite{BP}, Eq.~(2.14), p.~131)
\beq
\f=\int\frac{c\;dr}{r^2\,\sqrt{2\left(h-V(r)\right)-c^2/r^2}} +const\,,
		     \label{geosw71}
\eeq
where $c$ is the angular momentum per unit mass, one immediately identifies
$A_3$ with the angular momentum $c$. From (\ref{varsep2}), one has that in
the general case $\th\neq \pi/2$ the constant $A_2$ represents the component
of the angular momentum along the polar axis.

We can also obtain the physical meaning of the constant $A_3$ and the second
Kepler law from (\ref{tempro}) and (\ref{geosw4}). Assuming ``direct orbits"
we have:
\beq
\frac{d\f}{d\t} \equiv \frac{d\f}{dr}\;\frac{dr}{d\t}=\frac{A_3}{r^2}
\,.
\label{Kep}
\eeq
As for the constant $A_1$, in our units ($c=1, m=1$), $A_1^2-1$ must be identified
with twice the kinetic energy through the standard relativistic relations
$E^2-m^2c^4=c^2\,p^2$ and $p^2=2\,m\,T$, where $T$ is the classical kinetic energy.
Thus in our
units $A_1^2=E^2$; a further detailed investigation leads to $A_1=-E$.

Regarding the advance of the perihelion of Mercury, it is convenient to start
from (\ref{geosw5}) rewritten in terms of the variable $u=1/r$
\beq
\left(\frac{du}{d\f}\right)^2
=\frac{1}{A_3^2}\left[A_1^2-(1+A_3^2\,u^2)(1-\a\,u)\right]
		     \label{geosw8}	   \,.
\eeq
By differentiating this with respect to $\f$, we obtain
\beq
\frac{d^2 u}{d\f^2}+u=\2\frac{\a}{A_3^2}+\frac{3}{2}\,\a\,u^2\,,
\eeq
which since $\a=2\,G\,M$ can be rewritten
\beq
\frac{d^2 u}{d\f^2}+u=\frac{G\,M}{A_3^2}(1+3\,A_3^2\,u^2)\,.
\eeq
A comparison with the classic Binet equation (see \cite{BP}, Eq.~(2.24))
shows that the second term on the right hand side represents the
relativistic correction which accounts for the advance of the perihelion.
The reader can find the details of the calculations in the well known
text by Bergmann \cite{Ber}.

As a final remark, we point out that, returning to the geodesic equations
in the form of (\ref{geosw1}, \ref{geosw2}, \ref{geosw3}),
we can further obtain an  expression for $r$ as a function of $\f$ more
general than Eq.~(\ref{geosw4}). In fact, without the condition $\th=\pi/2$,
we have from Eq.~(\ref{geosw2}), $\th$ as a function of $\f$ and then,
from Eq.~(\ref{geosw3}), $r$ as a function of $\f$. This leads to
\beq
\cot\th=-\frac{\sin(\f-B_2)}{\sinh\e}
\label{geotefi}
\eeq
and
\beq
\sin[f(r)+B_3]=\frac{\cosh\e\,\sin(\f-B_2)}{\sqrt{\sinh^2\e+\sin^2(\f-B_2)}}
\,,
 \label{georrfi}
\eeq
where
$$f(r)=A_3\int\frac{dr}{\sqrt{A_1^2r^4-r(r^2+A_3^2)(r-\a)}}
$$
is the
elliptic integral in $dr$  of Eq.~(\ref{geosw3}). Eqs.~(\ref{geotefi}) and
(\ref{georrfi}) give the geodesics in the space as functions of the parameter $\f$.

Note that above we have obtained the result that the orbits are planar
without invoking the spherical symmetry of the field and without
setting $\th=\pi/2$.

\subsection{The Kerr metric}

The method based on Beltrami's theorem that we have applied so far to study
geodesic motion in the \sw \ spacetime can clearly be applied to the Kerr
spacetime as well. We start from the Kerr metric \cite{dino}
\beq
d\,s^2=\frac{\r^2\,\D}{\Si^2}d\,t^2-\frac{\Si^2}{\r^2}\left(d\,\f-\frac{2\,a\,M\,r}
{\Si^2}d\,t\right)^2\sin^2\th-\frac{\r^2}{\D}d\,r^2-\r^2\,d\,\th^2
\label{ggKe}\,,
\eeq
where $\r^2=r^2+a^2\,\cos^2\th,\,\D=r^2+a^2-2\,M\,r,\,\Si^2=(r^2+a^2)^2-
a^2\,\D\,\sin^2\th$; $M$ and $a$ are constants that in the Newton limit
represent the mass and the angular momentum per unit mass. \\
As is well known (see \cite{dino}, p.~289) the Kerr metric reduces to the
\sw \ metric when the constant $a=0$.

The contravariant form of the metric tensor is
\beq
g^{ij}=\frac{1}{\r^2}\left(\begin{array}{cccc}
\Si^2/\D& 0 &0&\,2\,a\,M\,r/\D \\
0&-\D& 0&0 \\
0&0&-1&0\\
2\,a\,M\,r/\D& 0&0&(a^2\,\sin^2\th-\D)/\D\,\sin^2\th		 \label{gKe}
\end{array}\right)			      \,,
\eeq
where the entries are arranged following the sequence $dt,dr,d\th,d\f$,
respectively.
Eq.~(\ref{B}) with the contravariant components of the metric tensor given by (\ref{gKe})
turns out to be
\beq
\frac{\Si^2}{\D}\left(\frac{\partial\, \t}{\partial \,t}\right)^2+
\frac{4\,a\,M\,r}{\D}\left(\frac{\partial\, \t}{\partial \,t}\right)
\left(\frac{\partial\, \t}{\partial \,\f}\right)-
\frac{\D-a^2\,\sin^2\th}{\D\,\sin^2\th}\left(\frac{\partial\, \t}{\partial \,
\f}\right)^2-\D\left(\frac{\partial\, \t}{\partial \,r}\right)^2-
\left(\frac{\partial\, \t}{\partial \,\th}\right)^2=\r^2\,.
\label{Ke}
\eeq
For (\ref{ggKe}), as for (\ref{Sw}), the coefficients of the line
element do not
depend on $t$ and $\f$, therefore we can seek a solution of (\ref{B})
in the same form of \sw \ equation
\beq
\t=A_1\,t+A_2\,\f+\t_1(r)+\t_2(\th) \label{tauk}\,.
\eeq
By substituting into Eq.~(\ref{Ke}) we obtain
\beq
\frac{1}{\D}\left[A_1\,(r^2+a^2)+A_2\,a\right]^2-
\frac{1}{\sin^2\th}\left[A_1\,a\,\sin^2\th+A_2\right]^2-\D\,
\left(\frac{d\, \t_1}{d \,r}\right)^2-
\left(\frac{d\, \t_2}{d \,\th}\right)^2= r^2+a^2\,\cos^2\th\,,
\eeq
and
\beq
\frac{1}{\D}\left[A_1\,(r^2+a^2)+A_2\,a\right]^2-r^2-\D\,
\left(\frac{d\, \t_1}{d\,r}\right)^2=
\frac{1}{\sin^2\th}\left[A_1\,a\,\sin^2\th+A_2\right]^2
+\left(\frac{d\, \t_2}{d\,\th}\right)^2+a^2\,\cos^2\th\,.
\eeq
The left hand side is a function of $r$, the right hand side is a positive function
of $\th$; then introducing a new separation constant $A_3^2$ we obtain the solutions
\ba
\t_1=\int\frac{\sqrt{[A_1\,(r^2+a^2)+A_2\,a]^2-\D(r^2+A_3^2)}}{\D}\,d\,r, \\
\t_2=\int\frac{
\sqrt{(A_3^2-a^2\,\cos^2\th)\sin^2\th-(A_1\,a\,\sin^2\th+A_2)^2}}{\sin\th}\,d\,\th\,.
\ea
If we set
\beq
R(r)=[A_1\,(r^2+a^2)+A_2\,a]^2-\D(r^2+A_3^2),\quad
\Th(\th)=(A_3^2-a^2\,\cos^2\th)\sin^2\th-(A_1\,a\,\sin^2\th+A_2)^2\,,
\eeq
Eq.~(\ref{tauk}) becomes
\beq
\t=A_1\,t+A_2\,\f+\int^r\frac{\sqrt{R(r)}}{\D}\,d\,r+\int^\th \frac{
\sqrt{\Th(\th)}}{\sin \th}
\,d\,\th
\label{tauk1}\,.
\eeq

The equations for the geodesics can now be obtained by the standard procedure
of Eq.~(\ref{t1}) following from Beltrami's theorem. We get
\ba
\frac{\partial \t}{\partial A_1}\equiv t+ \int\frac{(r^2+a^2)
[A_1(r^2+a^2)+A_2\,a]}{\D\,\sqrt{R}}\,dr- \int\frac{(A_1\,a\,\sin^2\th+A_2)
a\sin\th}{\sqrt{\Th}}d\,\th  =B_1\,,
\label{geoke1}\\
\frac{\partial \t}{\partial A_2}\equiv \f+
\int a\,\frac{A_1(r^2+a^2)+A_2\,a}{\D\,\sqrt{R}}\,dr-
\int\frac{(A_1\,a\,\sin^2\th+A_2)}{\sin\th\sqrt{\Th}}d\,\th=B_2\,,
\label{geoke2} \\
\frac{\partial \t}{\partial A_3}\equiv A_3\left[-\int\frac{dr}{
\sqrt{R}} +\int\frac{\sin\th}{\sqrt{
\Th}}\, d\th \right]=B_3  \,.
\label{geoke3}
\ea
Moreover, we can calculate the proper time with the same procedure we have used for
the \sw \ metric. We start from the expression (\ref{tau5}) and obtain
\be
\t = const -\int \frac{r^2}{\sqrt{R}}\,dr-\int\frac{a^2\,\cos^2\th\,
\sin\th}{\sqrt{\Th}}\;d\th \,. \label{tauke2}
\ee
Now the integral in $d\th$ does not vanish, while from ({\ref{geoke3}) we
can find $\th$ as a function of $r$ and then find $\t(r)$. In any case from
Eqs.~(\ref{geoke2}, \ref{geoke3}, and \ref{tauke2}) we can obtain the relations
analogous to those obtained for the \sw \ metric.

\section{Conclusions}

It turns out, as Beltrami himself was the first to point out, that Beltrami's method is
formally analogous to the method of integration of the Hamilton-Jacobi equation.
On the other hand this is even more evident if we consider that the differential
parameter $\D_1\,\t$ is formally analogous to the expression
$${\cal H}=g^{i\,j}\,p_i\,p_j$$
which represents the Hamiltonian and then
$\D_1\,\t=1$ is equivalent to the Hamilton-Jacobi equation.

However, in spite of this formal analogy, by using Beltrami's theorem we remain in
a geometric context to obtain the geodesics, i.e., the orbits of a test
particle in the gravitational field, without being obliged to resort to concepts
copied from classical mechanics. In fact it must be remarked that very often
one uses too freely in general relativity procedures which obtain a true legitimacy
only for $r\ra \infty$ or weak gravitational fields. The fact that, a posteriori,
things turn out to be correct in the nonrelativistic limit does not always remove the ambiguity from
certain formulations.

Finally we note that the ``geometrical
integration'' here described  allows us to recover the classical conservation laws
instead of introducing them a priori.

\subsection*{Acknowledgments}
We are grateful to Robert Jantzen for fruitful discussions.



\end{document}